# One-step, Wash-free, Nanoparticle Clustering-based Magnetic Particle Spectroscopy (MPS) Bioassay Method for Detection of SARS-CoV-2 Spike and Nucleocapsid Proteins in Liquid Phase

Kai Wu[a,†,*], Vinit Kumar Chugh[a,†], Venkatramana D. Krishna[b,†], Arturo di Girolamo[a], Yongqiang Andrew Wang[c], Renata Saha[a], Shuang Liang[d], Maxim C-J Cheeran[b,*], and Jian-Ping Wang[a,*]

[a]Department of Electrical and Computer Engineering, University of Minnesota, Minneapolis, MN 55455, United States

[b]Department of Veterinary Population Medicine, University of Minnesota, St. Paul, MN 55108, United States

[c]Ocean Nano Tech LLC, San Diego, CA 92126, United States

[d]Department of Chemical Engineering and Material Science, University of Minnesota, Minneapolis, MN 55455, United States

**ABSTRACT:** With the ongoing global pandemic of coronavirus disease 2019 (COVID-19), there is an increasing quest for more accessible, easy-to-use, rapid, inexpensive, and high accuracy diagnostic tools. Traditional disease diagnostic methods such as qRT-PCR (quantitative reverse transcription-PCR) and ELISA (enzyme-linked immunosorbent assay) require multiple steps, trained technicians, and long turnaround time that may worsen the disease surveillance and pandemic control. In sight of this situation, a rapid, one-step, easy-to-use, and high accuracy diagnostic platform will be valuable for future epidemic control especially for regions with scarce medical resources. Herein, we report a magnetic particle spectroscopy (MPS) platform for detection of SARS-CoV-2 biomarkers: spike and nucleocapsid proteins. This technique monitors the dynamic magnetic responses of magnetic nanoparticles (MNPs) and uses their higher harmonics as a measure of the nanoparticles' binding states. By anchoring polyclonal antibodies (pAbs) onto MNP surfaces, these nanoparticles function as nanoprobes to specifically bind to target analytes (SARS-CoV-2 spike and nucleocapsid proteins in this work) and form nanoparticle clusters. This binding event causes detectable changes in higher harmonics and allows for quantitative and qualitative detection of target analytes in liquid phase. We have achieved detection limits of 1.56 nM (equivalent to 125 fmole) and 12.5 nM (equivalent to 1 pmole) for detecting SARS-CoV-2 spike and nucleocapsid proteins, respectively. This MPS platform combined with one-step, wash-free, nanoparticle clustering-based assay method is intrinsically versatile and allows for the detection of a variety of other disease biomarkers by simply changing the surface functional groups on MNPs.

**KEYWORDS:** *COVID-19, SARS-CoV-2, spike protein, nucleocapsid protein, magnetic nanoparticle, magnetic particle spectroscopy, disease diagnostics*

## 1. INTRODUCTION

Severe acute respiratory syndrome coronavirus 2 (SARS-CoV-2), a novel human corona virus is the causative agent of ongoing coronavirus disease 2019 (COVID-19) pandemic. It was first identified in patients with acute respiratory illness in Wuhan, China in December 2019 and soon after became a public health emergency of international concern.[1] Since its emergence, SARS-CoV-2 infected 167 million people worldwide and caused 3.48 million deaths (as of May 25[th], 2021). The clinical presentation of SARS-CoV-2 infection ranges from asymptomatic or mild flu-like symptoms to fatal pneumonia and acute respiratory illness.[2] The SARS-CoV-2 belongs to the genus betacoronavirus of the family Coronaviridae. They are enveloped, single-stranded, positive sense RNA viruses with a genome size of approximately 29.9 kb.[3–5] The SARS-CoV-2 is about 60-140 nm in diameter and consists of four structural proteins and sixteen non-structural proteins.[6] The structural proteins such as spike (S), membrane (M), and envelope (E) are associated with viral envelope and nucleocapsid (N) protein forms the capsid outside genomic RNA.

Early diagnosis is critical for clinical management of patients and controlling the spread of COVID-19 by isolating infected patients. Several nucleic acid detection tests targeting N, E, S or RNA-dependent RNA polymerase (RdRp) genes of the virus are commercially available for the diagnosis of COVID-19.[7–9] Viral RNA detection by quantitative reverse transcription polymerase chain reaction (RT-qPCR) is currently the gold standard test for SARS-CoV-2 infection.[10,11] Although this nucleic acid amplification test is highly sensitive with the limit of detection as low as 1 RNA copy/µL, RT-PCR needs expensive laboratory instrument, skilled technicians, and days to get the result. Therefore, there is an urgent need for a simple, rapid, sensitive, and accurate point-of-care (POC) diagnostic platform that can detect SARS-CoV-2 directly from clinical samples. Antigen detection test can directly detect SARS-CoV-2 in respiratory specimens by detecting virus specific antigens.[12–14] However, some of the commercially available antigen detection tests have very low sensitivity.[15]

Magnetic particle spectroscopy (MPS) is one of the most promising candidates for rapid, inexpensive, and high accuracy bioassays.[16–22] It is a homogeneous biosensing tool that monitors the dynamic magnetic responses of magnetic nanoparticles (MNPs) in liquid phase.[23–26] By applying AC magnetic fields to MNPs, the magnetic moments of MNPs follow the time-varying external field directions through a combined Néel and Brownian relaxation mechanism.[16,18,27–29] The dominant relaxation of these two processes is dependent on the magnetic core sizes of MNPs (assuming unconstrained MNPs in liquid without surface binding of any chemical substances).[30–32] The Brownian relaxation process is dominant for single-core iron oxide MNPs with core sizes above 20 nm (this critical size may vary for different magnetic materials of different magnetic properties).[33–37] This relaxation process reflects the degree of freedom of physical rotational motion of MNPs. Conjugation of any chemical substances including protein molecules, aptamers, and other non-magnetic materials onto MNPs can hinder or even block this Brownian relaxation.[23,27,38–45] As a result, this binding event causes a phase lag between magnetic moments and external AC fields (except at very low frequencies) and thus, weaker dynamic magnetic responses are observed from clustered MNPs. By exploiting this unique property of Brownian relaxation-dominant MNPs, researchers have applied it along with MPS platform for the detection of H1N1 virus[39], thrombin[38], staphylococcal toxins[20], SARS-CoV-2[23,46], botulinum neurotoxins A, B, and E[21], etc. Compared to the traditional optical, mechanical, and electrochemical sensing techniques, this magnetic sensing technique is immune to the background noise from biological samples that may interfere with the signal reading since most biological substrates are non-magnetic (or paramagnetic) and MNPs are the only sources of magnetic signal.

Our group has successfully built up both benchtop and portable MPS systems for one-step, wash-free bioassays.[22,39] In this paper, we have modified the MPS system to include a one-stage lock-in scheme for the advantages of down sampling and phase sensitive rejection of noise. Herein, we report a one-step, wash-free, nanoparticle clustering-based MPS bioassay method for the detection of SARS-CoV-2 spike and nucleocapsid proteins in liquid phase using this one-stage lock-in MPS system. By surface conjugating polyclonal antibodies (pAbs) to MNPs, each MNP will be able to specifically bind to targe protein molecules. On the other hand, each protein molecule has multiple epitopes that allows the specific bindings of pAbs. Thus, the pAbs anchored to MNP surfaces and the nature of different epitopes from each protein molecule allows us to amplify the binding events caused dynamic magnetic responses change. We demonstrate that this nanoparticle clustering-based detection method can detect target analytes by monitoring the dynamic magnetic responses of MNPs.

## 2. MATERIALS AND METHODS

**2.1. Materials.** The SARS-CoV-2 recombinant nucleocapsid protein (His tag, consists of 430 amino acids and predicts a molecular mass of 47.08 kDa, Cat: 40588-V08B), the recombinant spike protein (RBD, His tag, consists of 234 amino acids with a molecular mass of 26.54 kDa, Cat: 40592-V08H), the nucleocapsid antibody (polyclonal rabbit IgG, Cat: 40588-T62), and the spike RBD antibody (polyclonal rabbit IgG, Cat: 40592-T62) are purchased from Sino Biological Inc. The IPG30 MNPs are 30 nm iron oxide nanoparticles functionalized with protein G, with weight concentration of 1.7 mg/mL and particle concentration of 57.8 nM, provided by Ocean NanoTech. The phosphate-buffered saline (PBS, Cat: 79378) is purchased from Sigma-Aldrich Inc.

**2.2. Magnetic Property Characterization.** 10 μL IPG30 MNP suspensions are pipetted onto a filter paper and air-dried. Then the static magnetic hysteresis loops are collected on a Physical Properties Measurement System (PPMS, Quantum Design Inc.), to obtain the magnetic properties of these nanoparticles such as the saturation ($M_s$) and coercivity ($H_c$). The static magnetic hysteresis loops of IPG30 MNPs are measured under external fields of -5000 – 5000 Oe and -500 – 500 Oe, and the magnetic properties are analyzed in the Supporting Information S1.

**2.3. One-stage Lock-in Magnetic Particle Spectroscopy (MPS) System.** The MPS platform used in this work utilizes frequency mixing approach where the excitation magnetic field consists of a low-frequency ($f_L$) component and a high-frequency ($f_H$) component.[17,19–22,26,39,43] A low frequency field having 50 Hz frequency, 250 Oe amplitude and a high-frequency field having 5000 Hz frequency, 25 Oe amplitude were used for the magnetic excitation of the MNPs. In this work, the MPS system is modified with the addition of a one-stage lock-in implementation to improve the detection sensitivity of the system. Figure 1(a) shows the schematic representation of the sensing scheme used for decoding the MNP response signal in the one-stage lock-in MPS system. The signal from pick-up coils is amplified using high-precision instrumentation amplifier, INA828 from Texas Instruments. Figures 1(b) & (c) show the time-domain and frequency-domain representations of the MNP signal at this stage. The amplified signal is then processed by a one-stage lock-in implementation consisting of a synchronous demodulator followed by band-pass filtering. AD630 from Analog Devices is used for the synchronous demodulator application. Figures 1(d) & (e) depict the corresponding time-domain and frequency-domain presentations of MNP signal after the lock-in implementation stage. This one-stage lock-in MPS system helps remove the feedthrough signals corresponding to the excitation magnetic field frequencies and only records the dynamic magnetic responses of MNPs, hence improving the sensitivity. The filtered signal is sampled at 100 KSPS sampling rate and communicated to a laptop for postprocessing and data analysis. Each MPS reading for a vial contains 170,000 samples which are collected over a period of roughly 2 seconds. If multiple readings are required from a sample, we use a cooldown time of 3 minutes between the readings to ensure neither the coils nor the analog driving circuitry heats up too much.

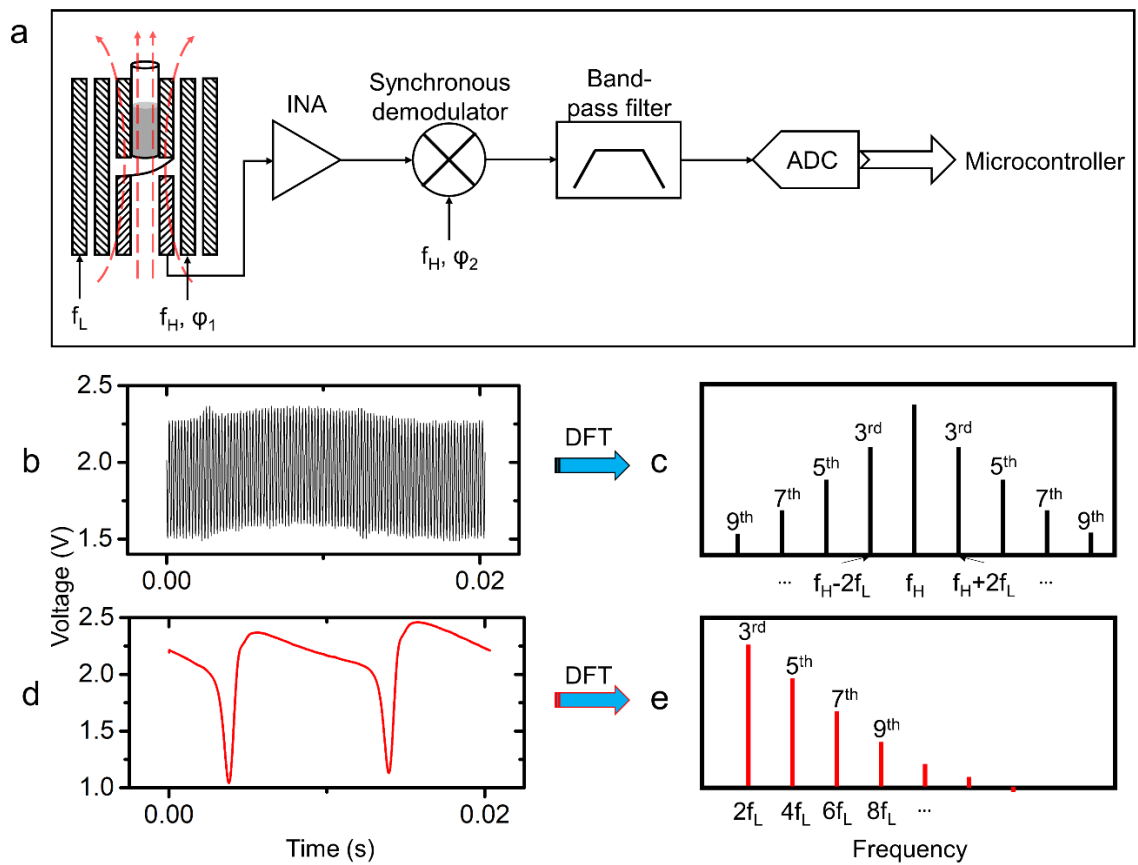

Figure 1. (a) Schematic diagram of the MPS system with a one-stage lock-in implementation. (b) is the time-domain voltage signal collected from pick-up coils after INA amplification stage and (d) shows the voltage signal post lock-in implementation. (c) and (e) are the MPS spectra from (b) and (d) after discrete Fourier transform (DFT), respectively.

**2.4. MPS-based Bioassay Method.** The MPS-based bioassays for detection of target analytes in liquid phase allows for one-step, wash-free assays that can be handled by a layperson without much technical training requirements. This method relies on the intentional interference on the Brownian relaxation of MNPs in liquid. When subjected to external AC magnetic fields (excitation fields), the magnetic moments of MNPs try to re-align to external field directions though the Brownian relaxation. This process will be hindered if MNPs are surface bound with polymers, protein molecules (such as antibody, antigen), aptamers, etc.[18,23,26,38], fixed on a surface/substrate[20,21,46,47], or form nanoparticle clusters[26,38,39]. Figure 2(b) shows a schematic view that by surface coating an increased number of antibodies, the dynamic magnetic responses of MNPs (in the form of MPS spectra, i.e., the amplitudes of higher harmonics) become weaker. In the presence of target analytes, the antibody functionalized MNPs bind to these analytes and form nanoparticle clusters. As a result, this nanoparticle clustering events further weaken the dynamic magnetic response and thus, lowering the harmonic amplitudes, as shown in Figure 2(c). On the other hand, if less target analytes are present, then less nanoparticle clusters are formed and as a result, the dynamic magnetic responses of MNPs yield stronger harmonics (Figure 2(d)).

**2.5. Experimental and Control Groups.** We firstly evaluated the detection limit of our one-stage lock-in MPS system for the detection of IPG30 MNPs. Samples containing a varying concentration/amount of IPG30 MNPs are prepared by two-fold dilutions as shown in Table S1 from Supporting Information S2. Three independent MPS readings are taken from each sample and the amplitudes of higher harmonics are summarized in Figure S2 from Supporting Information S2. Results show that the developed one-stage lock-in topology-based MPS system can detect as low as 266 ng (equal to 9 fmole) of IPG30 MNPs, which equals to 512-fold dilutions of the original concentration. Compared with our previous work, this modified MPS system has higher detection sensitivity.[22,26,39] Thus, for the bioassays in this work, we used diluted IPG30 MNPs with final particle concentration of 3.06 nM and weight concentration of 90 µg/mL. This feasibility attempt allows us to explore inexpensive bioassays on our MPS platform for future high-volume tests at the users' ends as well as in regions with scarce medical resources. Furthermore, it is reported that for MPS-based homogeneous bioassays, lower MNP concentrations could improve the detection sensitivities of biomolecules.[24,48]

The design of one-step, wash-free, nanoparticle clustering-based MPS bioassays is highly relied on the surface functionalization of polyclonal antibodies (pAbs) on MNPs. It has been reported in our previous work that the pAbs functionalized on MNPs can trigger the cross-linking between MNPs and target analytes thus, forming MNP clusters.[39] In this work, the SARS-CoV-2 spike and nucleocapsid proteins have multiple epitopes that can be recognized by their corresponding pAbs. As a result, each protein molecule (i.e., spike and nucleocapsid) can bind to multiple pAbs from multiple MNPs. On the other hand, each MNP functionalized with multiple pAbs can bind to multiple protein molecules. A schematic example is given in Figure 2(c).

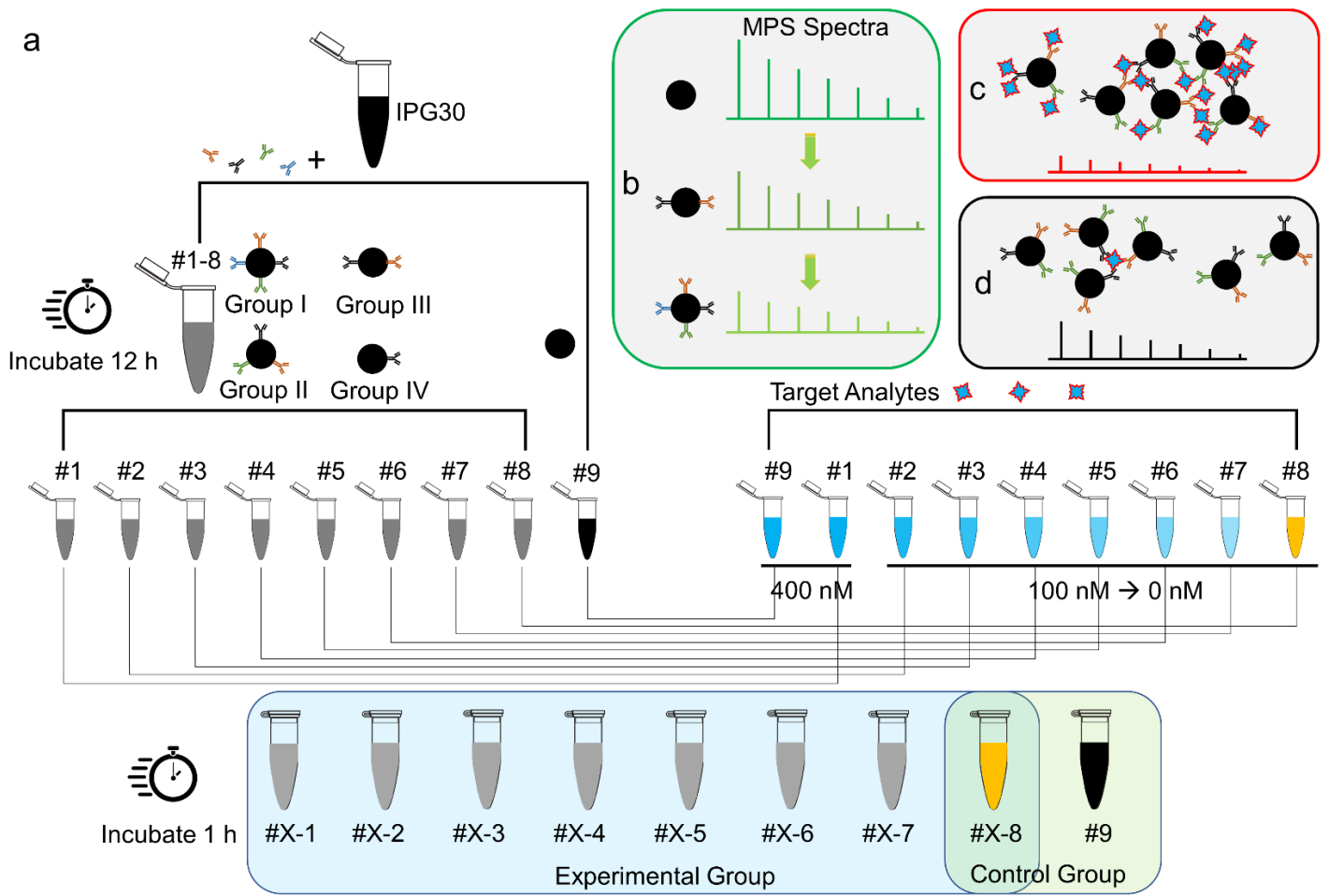

Figure 2. (a) Schematic view of samples prepared for one-step, wash-free, nanoparticle clustering-based MPS bioassays. For spike and nucleocapsid protein tests, four groups of samples are prepared by varying the amount of pAbs functionalized on MNPs. Group I: MNP:pAb=1:4, Group II: MNP:pAb=1:3, Group III: MNP:pAb=1:2, and Group IV: MNP:PAb=1:1. Vial #9 is a negative control sample with bare MNPs (no pAbs functionalized). Spike and nucleocapsid proteins are prepared at varying concentrations from 400 nM to 0 nM. (b) compares the MPS spectral intensities of MNPs functionalized with different numbers of pAbs. (c) and (d) compare the MPS spectral intensities of surface functionalized MNPs in the presence of (c) high and (d) low concentrations/amounts of the target analytes.

As shown in Figure 2(a), four experiment groups are designed by varying the number of pAby functionalized on MNPs: 1) Group I, MNP:pAb=1:4 (i.e., in average, there are 4 pAbs functionalized on each MNP), 2) Group II, MNP:pAb=1:3, 3) Group III, MNP:pAb=1:2, and 4) Group IV, MNP:pAb=1:1. Each group consists of 8 surface functionalized IPG30 MNP samples added with varying concentration/amount of spike or nucleocapsid proteins, as listed in Table 1. For the SARS-CoV-2 spike protein tests, we prepared 4 groups of samples along with one negative control sample (vial #9) that contains bare IPG30 MNP without pAb coating (labeled as 'bare MNP' in this work). In each group, the vials #X-8 (X=I, II, III, and IV) serve as the negative control that contain

MNPs functionalized with varying amount of pAbs while 0 nM of protein (we replaced the protein with PBS buffer) is added. Thus, a total of 33 samples for SARS-CoV-2 spike protein and 33 samples for nucleocapsid protein tests are prepared, respectively.

As shown in Figure 2(a), the diluted IPG30 MNPs (particle concentration of 3.06 nM) are incubated with spike or nucleocapsid pAbs for 12 hours at 4 °C on incubating shaker. This step allows the Fc part of IgG pAbs to bind to protein G from IPG30 MNP surfaces. Then 80 μL of pAb functionalized MNPs is mixed with 80 μL of spike or nucleocapsid proteins and incubated at room temperature for 1 hour with continuous shaking. After this point, the specific binding between pAbs and target protein molecules reaches to an equilibrium state, sample vials are then stored at 4 °C before MPS tests. It should be noted that the 1-hour incubation time of pAb functionalized MNPs with protein biomarkers is not optimized. To ensure the antibody-antigen binding to fully saturate, we set an incubation time of 1 hour. Further works should be carried out to reduce the incubation time by means of changing incubation conditions (such as temperature[49,50], pressure[51], flow rate[52], etc) or choosing higher binding affinity pAbs[53].

Table 1. Experimental and Control Group Designs for MPS-based SARS-CoV-2 Spike (S) and Nucleocapsid (N) Protein Tests.

| Sample Index | IPG30 MNP Amount/Vial | S/N PAb Amount/Vial | MNP:pAb | S/N Protein Amount/Vial[1] |
|---|---|---|---|---|
| Group I Vial 1-8 | 226 fmole | 904 fmole | 1:4 | #X-1: 400 nM (32 pmole)[2] <br> #X-2: 100 nM (8 pmole) |
| Group II Vial #1-8 | 226 fmole | 678 fmole | 1:3 | #X-3: 25 nM (2 pmole) <br> #X-4: 12.5 nM (1 pmole) |
| Group III Vial #1-8 | 226 fmole | 452 fmole | 1:2 | #X-5: 6.25 nM (500 fmole) <br> #X-6: 3.13 nM (250 fmole) |
| Group IV Vial #1-8 | 226 fmole | 226 fmole | 1:1 | #X-7: 1.56 nM (125 fmole) <br> #X-8: 0 nM (0 fmole) |
| Vial #9 | 226 fmole | 0 fmole | NA | #9: 400 nM (32 pmole) |

[1] S and N stand for SARS-CoV-2 spike (S) and nucleocapsid (N) proteins, respectively.

[2] X=I, II, III, and IV. There are four groups of samples prepared for S and N protein tests, respectively.

## 3. RESULTS AND DISCUSSIONS

**3.1 Time- and Frequency-Domain Signals Recorded by One-stage Lock-in MPS System.** The real-time voltage signals of blank sample (no IPG30 MNPs loaded, i.e., noise floor), bare MNPs (vial #9, IPG30 MNPs without surface functionalization), vial #IV-8 (MNP:pAb=1:1), and vial #I-8 (MNP:pAb=1:4) as recorded from pick-up coils are plotted in Figure 3(a). The background signal from blank sample is a relatively smooth sinusoidal waveform while all samples with MNPs loaded show visible spikes. These spikes are caused by the dynamic magnetic responses of MNPs in liquid phase under external AC magnetic fields (excitation fields) as explained in Section 2.4. The bare MNPs cause relatively stronger spikes (red curve) than the MNPs functionalized with one pAb each (blue curve, MNP:pAb=1:1, vial #IV-8), and the MNPs functionalized with four pAbs each (magenta curve, MNP:pAb=1:4, vial #I-8) show the weakest spikes. A zoomed in view of these spikes is presented in Figure 3(b) and the amplitudes of these pikes are marked in dashed lines. The frequency domain MPS spectra are calculated by carrying out discrete Fourier transform (DFT) on the time domain voltage signal. As shown in Figure 3(c), the bare MNP sample shows highest harmonic amplitudes, followed by MNP:pAb=1:1 (vial #IV-8) and MNP:pAb=1:4 (vial #I-8), and the blank samples showing the weakest and negligible harmonics. The results confirm that conjugating pAbs on MNP surfaces does hinder the physical rotational motion of MNPs and causes weaker harmonics. In addition, based on the extent of harmonic signal drop (compared with bare MNP sample) from MNP:pAb=1:1 and MNP:pAb=1:4, we can confirm that the pAbs are successfully functionalized onto MNPs. The functionalization of different amount of pAbs on MNPs is further confirmed by the hydrodynamic sizes (see Supporting Information S3). In addition, the colloidal stability of MNPs after coating pAbs is confirmed by zeta potential results as summarized in Supporting Information S3.

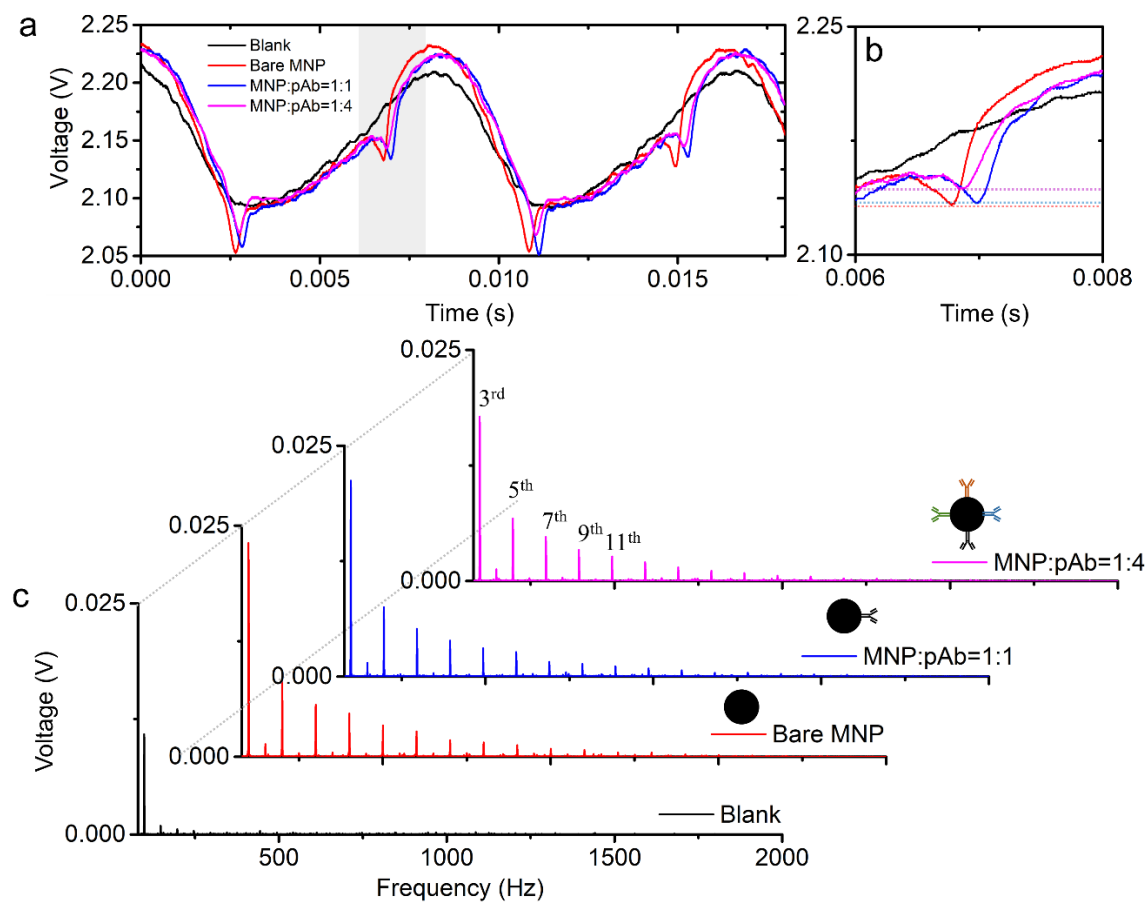

Figure 3. (a) Comparison of the time-domain voltage signal from blank sample (floor signal), bare MNP, MNPs each coated with one pAb, and MNPs each coated with four pAbs. (b) is the zoom in view of how the presence of MNP as well as the surface functionalization of pAbs affect the real-time voltage signal. (c) is the corresponding MPS spectra from (a). Higher harmonics such as the $3^{rd}$, the $5^{th}$, the $7^{th}$, the $9^{th}$, the $11^{th}$ harmonics are labeled in (c). Dotted lines have been added to help guide the view to compare the relative intensities of harmonics between the four sets of data.

**3.2. Concentration-Response Profiles of SARS-CoV-2 Spike Protein.** Six independent MPS readings are taken from each of the 33 samples and the $3^{rd}$ harmonics from 32 samples (Group I – IV) are summarized in Figure 4(a – d). Two-sample t-test is applied to compare the differences between two sets of harmonic data collected from different samples.

For the scenario of coating four pAbs onto each MNP (Group I, MNP:pAb=1:4, Figure 4(a)), adding 400 nM (equivalent to 32 pmole) of spike protein to vial #I-1 causes the $3^{rd}$ harmonic to be significantly lower (p<0.001) than the $3^{rd}$ harmonics from other vials in Group I. This weaker harmonic signal is due to the abundancy of target spike protein molecules from the sample that causes the nanoparticle clustering, as explained in Section 2.4 and Figure 2(c) & (d). However, vials #I-2 to #I-8 show no significant differences.

In Group II (MNP:pAb=1:3, Figure 4(b)), vials #II-1 to #II-4 show similar harmonic amplitudes and are significantly lower than other vials from the same group. Which indicates that in this scenario (MNP:pAb=1:3), adding 32 pmole (400 nM), 8 pmole (100 nM), 2 pmole (25 nM), and 1 pmole (12.5 nM) SARS-CoV-2 spike protein can completely block the Brownian relaxation of MNPs and cause the lowest achievable MPS spectra (or weakest dynamic magnetic responses). By reducing the number of spike protein in vials #II-5 to #II-8, the $3^{rd}$ harmonic amplitudes gradually increase due to less amount of target spike protein molecule present in the sample that can cause the nanoparticle clustering. From Group II, a clear linear concentration-response curve is observed from vials #II-4 to #II-8. Indicating that this bioassay design could be used for quantitative assays of spike protein samples within a concentration range of 0 – 12.5 nM (corresponding to spike protein amount of 0 – 1 pmole) and qualitative assays of spike protein samples with concentrations above 12.5 nM.

On the other hand, by functionalizing two pAbs (Group III, MNP:pAb=1:2, Figure 4(c)) and one pAb (Group IV, MNP:pAb=1:1, Figure 4(d)) per MNP, no significant response curves are observed. Although there is a clear trend showing that the harmonic signal amplitude decreases as we add a higher number of spike protein molecules.

Figure 4(e) compares the $3^{rd}$ harmonic amplitudes from vials #X-1 (X=I, II, III, and IV), bare MNP (vial #9), and blank (no IPG30 MNP, i.e., noise floor). All the vials added with IPG30 MNPs show significantly higher harmonic signal intensities than the blank sample. Of which the bare MNP (vial #9) sample shows the highest harmonic signal since no pAbs are functionalized and the Brownian relaxation of MNPs is not hindered even by adding 400 nM (32 pmole) spike protein to it. Which confirms that MNPs without pAb functionalization will not form clusters in the presence of target analytes. In addition, the 3rd harmonics from vials #X-3, #X-5, #X-7, and #X-8 (X=I, II, III, and IV) are compared in Figure 4(f – i), respectively. A clear trend is observed that by adding the same amount of spike protein molecules, MNPs that functionalized with more pAbs show lower harmonic signal. Figure 4(i) proves the schematic drawing in Figure 2(b) that functionalizing more pAbs per MNP will cause weaker dynamic magnetic responses and lower harmonic signals. Which also confirms that the pAbs have been successfully functionalized onto IPG30 MNPs.

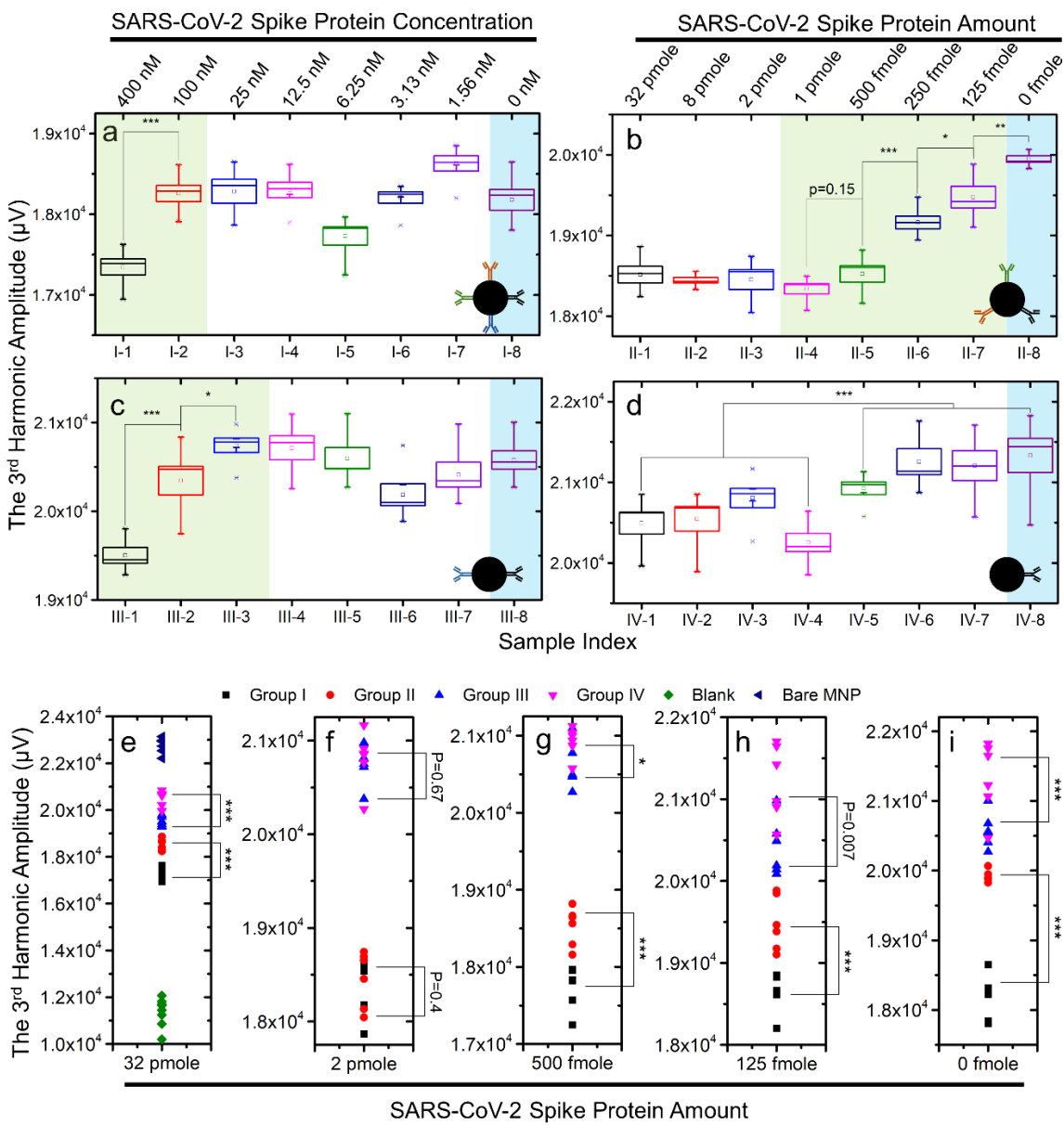

Figure 4. Concentration-response profiles of SARS-CoV-2 spike protein. (a) – (d) are the box plots of the 3rd harmonic amplitudes from vials #X-1 to #X-8, where X=I, II, III, and IV, respectively. The highlighted green areas are the monotonic concentration-response regions, and the highlighted blue areas are the vials #X-8, namely, control samples without the addition of target analytes. The middle horizontal line represents the averaged harmonic amplitude from 6 independent MPS readings, the top and bottom horizontal lines represent standard

errors, and the asterisks represent highest and lowest harmonic amplitudes recorded among 6 independent readings. (e) – (i) compare the 3$^{rd}$ harmonic amplitudes from vials #X-1, #X-3, #X-5, #X-7, and #X-8, respectively. *** $p < 0.001$; ** $p < 0.01$; * $p < 0.05$.

From the concentration-response profiles of SARS-CoV-2 spike protein, it is concluded that functionalizing three pAbs per MNP (Group II, MNP:pAb=1:3, Figure 4(b)) gives us the best linear response curve with a detection limit of 1.56 nM (equivalent to 125 fmole of spike protein molecules).

In Figure 5, we have summarized the ratios of higher harmonics (from the 5$^{th}$ to the 15$^{th}$ harmonics) to the 3$^{rd}$ harmonics collected from each sample. For Groups I, III, and IV (Figure 5(a), (c), and (d)), the harmonic ratio curves of vials #X-1 to #X-8 (X=I, III, and IV) are tightly distributed with very narrow gaps or even overlapping. While the harmonic ratio curves from Group II vials #II-1 to #II-8 are sparsely distributed. The significant differences in harmonic ratio curves from vials added with different concentration/amount of spike protein molecules allow us to analyze and collect meaningful concentration-response profiles as shown in Figure 4(b).

The harmonic amplitude decreases as the harmonic index increases, namely, the harmonic amplitudes $A3 > A5 > A7 > A9 > A11 > A13 > A15$, where A represents <u>a</u>mplitude (A). It is observed that functionalizing different amount of pAbs per MNP also changes the decay rate of higher harmonics (with harmonic index), as reflected in the MPS spectra from Figure 3(c). The yellow double arrows in Figure 5(a – d) mark the decay rates from the 5$^{th}$ to the 3$^{rd}$ harmonics across groups I to IV. In average, the R53 (harmonic amplitude ratio of A5 over the A3) from Groups I – IV are 0.91, 0.85, 0.73, and 0.72, respectively. And the R73 (harmonic amplitude ratio of the A7 over the A3) from Groups I – IV are 0.65, 0.61, 0.52, and 0.51, respectively. Thus, conjugating more pAbs on each MNP will make the higher harmonics decay slower.

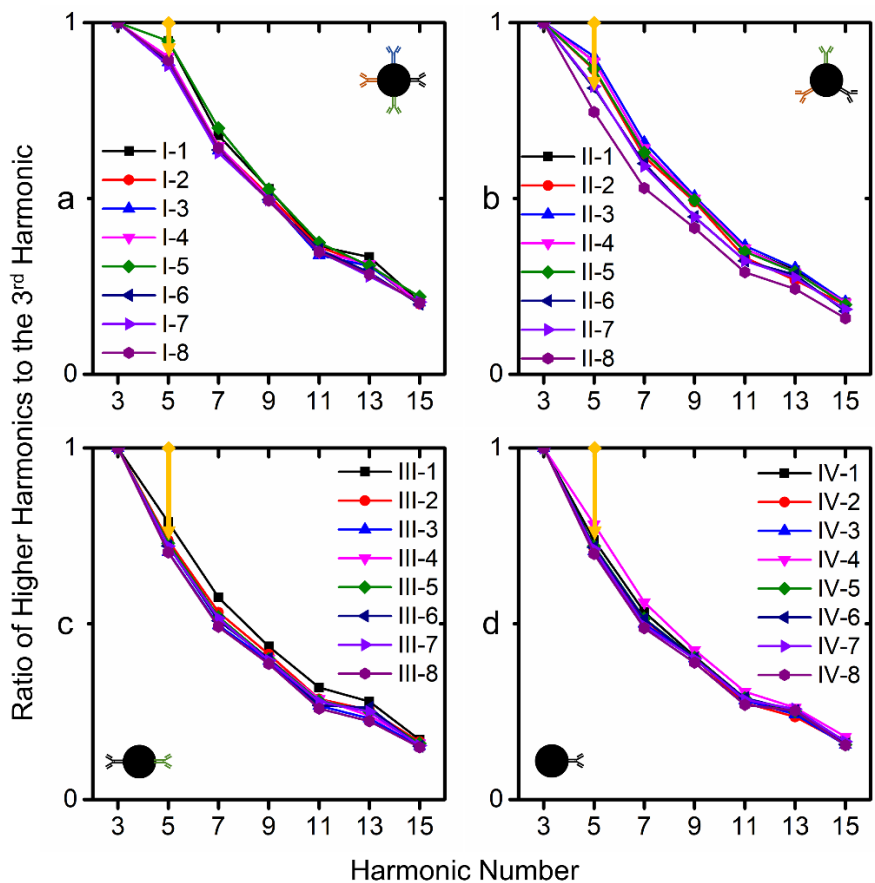

Figure 5. Ratios of higher harmonics to the 3rd harmonics recorded from Groups (a) I, (b) II, (c) III, and (d) IV. Yellow double arrows mark the decay rate from the 5th to the 3rd harmonics across groups I to IV.

Besides comparing each sample's higher harmonic amplitudes (A5 – A15) to its own 3rd harmonic (A3). We further compared the samples' higher harmonics from Group I – IV with the bare MNP sample (vial #9, IPG30 MNP without any surface functionalization). As shown in Supporting Information S4, the 3rd, 5th, 7th, 9th, 11th, 13th, and 15th harmonics of each sample is compared with the corresponding harmonics from bare MNP sample (vial #9) and the grayscale heatmaps of harmonic signal drop (defined as $\Delta = \frac{Ai_9 - Ai_{X-j}}{Ai_9} \times 100\%$, where $i$ is the harmonic index, the subscripts are sample indexes, X=I, II, III, and IV, $j$=1, 2, 3, …, 8) are plotted. Figure 6(a) compares the 3rd harmonic of each sample with the bare MNP sample (vial #9). In each row of Figure 6(a), by adding the same amount of spike protein molecules, the 3rd harmonic signal drop $\Delta$ decreases from I to IV, which agrees with the results in Figure 4(e – i). Thus, ideally, the color becomes darker from left column to right column in each grayscale heatmap (as schemed in Figure 6(b)). In each column of Figure 6(a), with IPG30 MNPs surface functionalized with an identical number of pAbs, adding more spike protein causes larger harmonic signal drop. Ideally, the color becomes darker from the bottom row (vials #X-1, X=I, II, III, and IV) to the top row (vials #X-8, X=I, II, III, and IV). Again, this ideal trend is schemed in Figure 6(b).

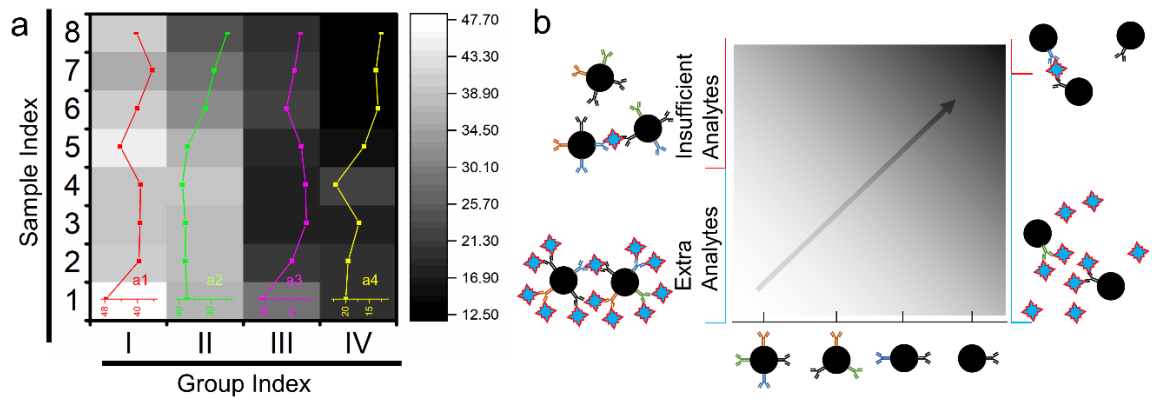

Figure 6. (a) Grayscale heatmaps of the 3$^{rd}$ harmonic signal drop Δ (in %) compared to bare MNPs for SARS-CoV-2 spike proteins. A total of 32 samples in Group I – IV are compared with the 3$^{rd}$ harmonic from bare MNPs (vial #9). (a1 – a4) are the harmonic signal drop, Δ, plotted as a function of spike protein amount/concentration for groups I – IV, respectively. (b) is the grayscale heatmap showing the ideal color trend regarding different amount of pAbs functionalized on MNPs as well as different scenarios of extra and insufficient target analytes (i.e., SARS-CoV-2 spike protein).

There are two variables in plotting the grayscale heatmaps of higher harmonic signal drop Δ: 1) the number of pAbs functionalized on each MNP (Group I – IV with MNP:pAb=1:4, 1:3, 1:2, and 1:1) and 2) the concentration/amount of spike protein added to each sample (i.e., vials #X-1 to #X-8, X=I, II, III, and IV). These two variables jointly cause the harmonic signal drop Δ increases from upper right corner to bottom left corner in diagonal. The 3$^{rd}$ harmonics give highest signal to noise ratio (SNR), and we can clearly see this trend from Figure 6(a). Figure 6(a1 – a4) plots the 3$^{rd}$ harmonic signal drop Δ curves for samples from Groups I – IV, where Group II shows the best monotonic concentration-response curve. Figure 6(b) schematically draws the scenarios where extra and insufficient target analytes (i.e., spike protein) are added. As a result, the number of target analytes directly affects the degree of nanoparticle clustering as well as the dynamic magnetic responses.

**3.3. Concentration-Response Profiles of SARS-CoV-2 Nucleocapsid Protein.** A total of 33 SARS-CoV-2 nucleocapsid samples are prepared in the same manner as the spike protein samples aforementioned. Six independent MPS readings are taken from each sample and the 3$^{rd}$ harmonics are plotted in Figure 7(a – d). Group I (MNP:pAb=1:4) shows remarkable monotonic concentration-response curve where the 3$^{rd}$ harmonics monotonically increase from vial #I-1 to vial #I-6, with nucleocapsid protein decreases from 400 nM (32 pmole) to 3.13 nM (500 fmole). Similar trends are also observed from Group II (MNP:pAb=1:3, Figure 7(b)) and Group III (MNP:pAb=1:2, Figure 7(c)) while both are less significant than Group I. The detection limit of SARS-CoV-2 nucleocapsid protein on our MPS system is 12.5 nM (equivalent to 1 pmole) based on the response curves from Group I.

We also observed that when functionalizing one pAb per MNP (Group IV), no remarkable response curves are observed from either spike or nucleocapsid proteins (Figure 4(d) and Figure 7(d)). For Group IV (MNP:pAb=1:1) in spike protein tests, vials #IV-1 to #IV-4 are significantly different ($p < 0.001$) from vials #IV-5 to #IV-8 but, no significant differences are observed within vials #IV-1 to #IV-4 or vials #IV-5 to #IV-8, as shown in Figure 4(d). Similar results are observed from Group IV (MNP:pAb=1:1) in nucleocapsid protein tests, vials #IV-1 to #IV-5 are significantly different ($p < 0.001$) from vials #IV-6 to #IV-8. Thus, for the detection of spike and nucleocapsid proteins, it's not practical to functionalize one pAb per MNP. The grayscale heatmaps of harmonic signal drop Δ from nucleocapsid samples are also provided in Supporting Information S5. The ratios of higher harmonics to the 3$^{rd}$ harmonics recorded from Groups I to IV for nucleocapsid protein can be found from Supporting Information S6.

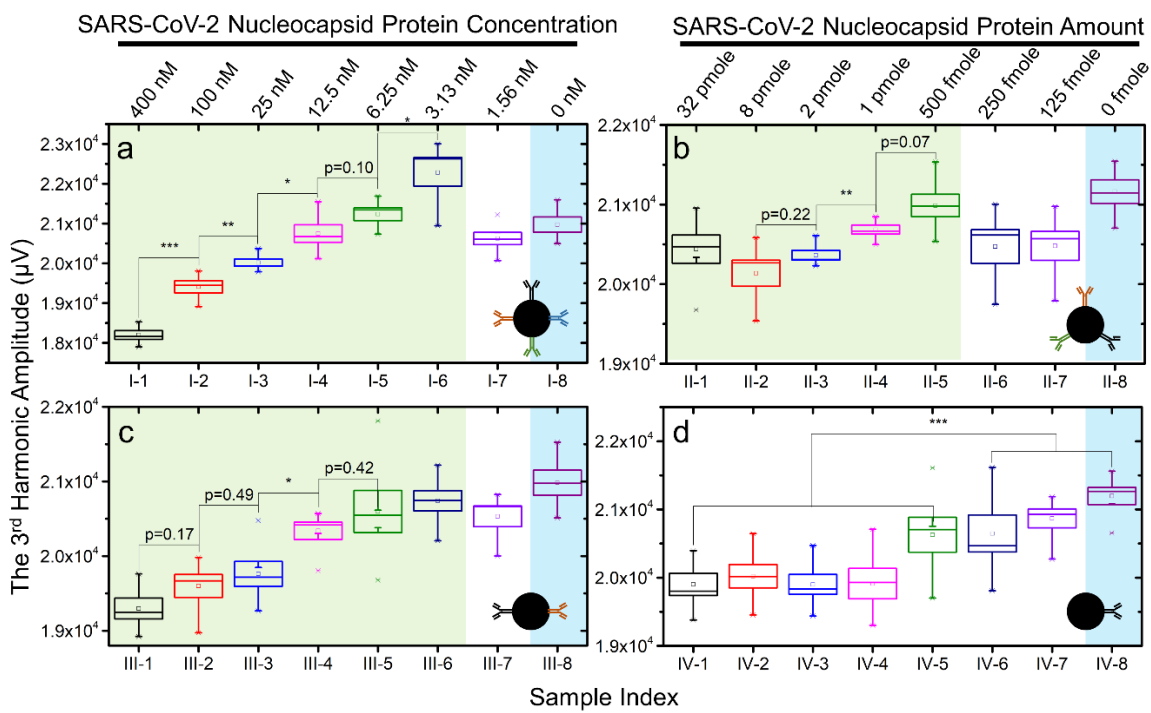

Figure 7. Concentration-response profiles of SARS-CoV-2 nucleocapsid protein. (a) – (d) are the box plots of the 3$^{rd}$ harmonic amplitudes from vials #X-1 to #X-8, where X=I, II, III, and IV, respectively. The highlighted green areas are the monotonic concentration-response regions, and the highlighted blue areas are the vials #X-8, namely, control samples without the addition of target analytes. Middle horizontal lines represent average signal, top and bottom horizontal lines represent standard errors. *** $p < 0.001$; ** $p < 0.01$; * $p < 0.05$.

## 4. CONCLUSIONS

Herein, we have successfully applied a one-stage lock-in MPS system for the detection of SARS-CoV-2 spike and nucleocapsid proteins directly in liquid phase. This one-step, wash-free detection method allows a layperson to carry out the tests without much technical training requirements. By conjugating pAbs to MNPs, it increases the chances of MNP binding events to target protein molecules (i.e., spike and nucleocapsid proteins in this work).

Combined with the nature that each protein molecule has different epitopes that allow the binding of multiple pAbs. We are able to amply the effect of MNP binding events caused obstruction of physical rotational motion of MNPs and thus, amplifying the changes in higher harmonics.

In addition, we explored the effect of pAbs amount functionalized on each MNP. Four groups of samples with MNP:pAb=1:4 (Group I), 1:3 (Group II), 1:2 (Group III), and 1:1 (Group IV) are designed. These pAb functionalized MNPs are used to quantitatively detect spike and nucleocapsid proteins varying from 0 fmole (vials #X-8) to 32 pmole (vials #X-1). The concentration-response profiles for spike and nucleocapsid proteins are generated based on the higher harmonics from MNPs. It is concluded that conjugating three pAbs per MNP yields a linear concentration-response curve and reaches a detection limit of 1.56 nM (equivalent to 125 fmole) for spike protein molecules. On the other hand, conjugating four pAb per MNP yield linear concentration-response curve for the detection of nucleocapsid proteins and a detection limit of 12.5 nM (equivalent to 1 pmole) is achieved. The higher harmonics from all the control groups: bare MNPs (vial #9) and vials #X-8 (X=I, II, III, and IV, where no proteins are added), confirmed that the pAbs have been successfully conjugated to MNPs.

To sum up, this nanoparticle clustering-based bioassay mechanism combined with MPS platform is intrinsically versatile and can be applied for the detections of other biomarkers such as proteins, viruses, nucleic acids, etc. Its nature of one-step wash-free experimental procedure allows for simpler and more convenient on field tests in the future. Zhong *et al.* also reported the SARS-CoV-2 spike protein detection from liquid phase using a similar MPS platform.[23] Where they coated anti-SARS-CoV-2 spike protein monoclonal antibodies (mAbs) to the MNPs and spike protein to 100 nm polystyrene (PS) beads. The binding events of MNPs to 100 nm PS beads through antibody-antigen specific recognition hinders the physical rotational motion of MNPs. Thus, in a similar manner, higher harmonics are recorded as indicators of spike protein abundancy. Each PS bead was coated with 100 SARS-CoV-2 spike proteins to mimic one SARS-CoV-2 virus particle. They demonstrated the detection limit of 84 fM (equal to 5.9 fmole) mimic virus particles, which is equal to 590 fmole of spike protein molecules. The main differences between our work and Zhong *et al.*'s work is that we utilized MNP-pAb-protein-pAb-MNP-pAb-protein-pAb-MNP-… clustering caused obstruction of physical rotational motion of MNPs to achieve specific detection of target SARS-CoV-2 spike and nucleocapsid protein. While Zhong *et al.* used MNP-mAb-protein-PS bead to block the physical rotational motion of MNPs. Although the detection strategy is different, the detection mechanism is the same, that is, relying on the change of the harmonic signal caused by the reduced rotational freedom of the MNPs to quantitatively and/or qualitatively detect the target SARS-CoV-2 protein biomarker.

It should also be noted that the results presented in this work are based on detecting purified spike and nucleocapsid proteins from PBS buffer. This is a feasibility demonstration of one-stage lock-in MPS system for protein biomarker detection, we did not measure SARS-CoV-2 biomarkers from clinical samples such as upper respiratory tract mucus including saliva and sputum. There are several factors that may impair the sensitivity and repeatability of MPS platforms when testing clinical samples, such as viscosity of biofluid samples, temperature, and nonspecific binding of unknown chemical compounds from biofluids. Since this one-step liquid phase detection mechanism relies on the binding events-caused Brownian relaxation change in MNPs, the viscosity and temperature of biofluid samples can cause different degrees of deviations in MPS harmonic signals, and the nonspecific binding can cause repeatability issues (as well as false positive results). To avoid the viscosity variation caused MPS signal bias, several samples preparation and collection methods can be adopted. Such as using exhaled breath condensate (EBC) that contains respiratory droplets from the lower respiratory tract.[54–56] The condensed water droplets containing SARS-CoV-2 virions can effectively avoid the body fluid (i.e., sputum, saliva) caused viscosity variations between patients. In addition to the sample collection method, the sample preparation method can also be altered for this specific application. Usually, the upper respiratory tract mucus samples (nasopharyngeal swabs, throat swabs, nasopharyngeal wash, saliva, sputum, etc) have high virus titer but the complex internal matrix could interfere the MPS test results due to the high viscosity and unknown chemical compounds that nonspecifically bind to MNPs.[57,58] The binding specificities of pAbs used in this work have been confirmed by ELISA testing (see Supporting Information S7). A mature sample preparation process has been applied to dilute specimens by washing buffers and filter out larger particle complexes.[59–61] To avoid the nonspecific binding of unknown chemical compounds from complex matrix, surface functionalized MNPs can be PEGylated[62–64]. Where polyethylene glycol (PEG) is a hydrophilic and neutrally charged polymer that can help prevent the nonspecific bindings of other chemical compounds to MNP surfaces.

## ASSOCIATED CONTENT

Supporting Information: S1. Static magnetic hysteresis loops and magnetic properties of IPG30 MNPs; S2. The minimum detectable amount of IPG30 by homebuilt MPS system; S3. Hydrodynamic size, zeta potential, the 3rd harmonic amplitudes of bare MNPs and MNPs functionalized with different amount of pAbs; S4. Grayscale heatmaps of higher harmonic signal drop Δ (in %) compared to bare MNPs for SARS-CoV-2 spike proteins; S5. Grayscale heatmaps of higher harmonic signal drop Δ (in %) compared to bare MNPs for SARS-CoV-2 nucleocapsid proteins; S6. The ratios of higher harmonics to the 3rd harmonics recorded from groups I to IV for SARS-CoV-2 nucleocapsid protein; S7. Specificity of SARS-CoV-2 spike and nucleocapsid pAbs confirmed by ELISA.

## AUTHOR INFORMATION

**Corresponding Authors**
*E-mail: wuxx0803@umn.edu (K. W.)
*E-mail: cheeran@umn.edu (M. C-J. C)
*E-mail: jpwang@umn.edu (J.-P. W.)

**Author Contribution**

K.W., M.C-J.C., and J.-P,W conceived the experiment. K.W. carried out partial of the MPS-based tests, plotted the figures, and wrote partial of the paper. V.K.C. carried partial of the MPS-based tests, designed the circuitry and assembled the circuit boards, and wrote partial of the paper. V.D.K. prepared all the experimental samples and wrote partial of the paper. A.G. assisted in the design of circuitry and assemble of circuit boards. Y.A.W. synthesized the MNPs. R.S. assisted in MPS raw data processing, the data collection of hydrodynamic size and zeta potential. S.L. assisted in MPS raw data processing and analysis. All authors proofread the paper.

**ORCID**

Kai Wu: 0000-0002-9444-6112

Vinit Kumar Chugh: 0000-0001-7818-7811

Venkatramana D. Krishna: 0000-0002-1980-5525

Arturo di Girolamo: 0000-0002-6906-8754

Yongqiang Andrew Wang: 0000-0003-2132-2490

Renata Saha: 0000-0002-0389-0083

Shuang Liang: 0000-0003-1491-2839

Maxim C-J Cheeran: 0000-0002-5331-4746

Jian-Ping Wang: 0000-0003-2815-6624

**Author Contributions**

[†]K.W., V.K.C., and V.D.K. have contributed equally to this work.

**Notes**

The authors declare no conflict of interest.

**ACKNOWLEDGMENTS**

This study was financially supported by the Institute of Engineering in Medicine, the Robert F. Hartmann Endowed Chair professorship, the University of Minnesota Medical School, and the University of Minnesota Physicians and Fairview Health Services through COVID-19 Rapid Response Grant. This study was also financially supported by the U.S. Department of Agriculture - National Institute of Food and Agriculture (NIFA) under Award Number 2020-67021-31956. Research reported in this publication was supported by the National Institute Of Dental & Craniofacial Research of the National Institutes of Health under Award Number R42DE030832. The content is solely the responsibility of the authors and does not necessarily represent the official views of the National Institutes of Health. Portions of this work were conducted in the Minnesota Nano Center, which is supported by the National Science Foundation through the National Nano Coordinated Infrastructure Network (NNCI) under Award Number ECCS-1542202.

**TOC:**

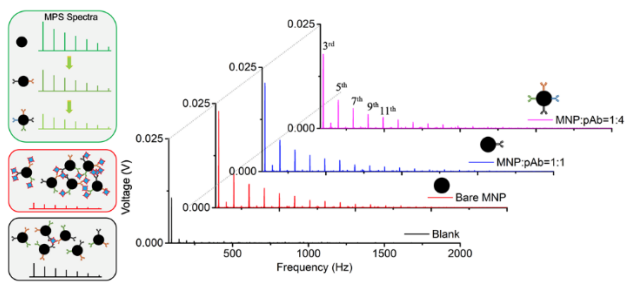

# Supporting Information

# One-step, Wash-free, Nanoparticle Clustering-based Magnetic Particle Spectroscopy (MPS) Bioassay Method for Detection of SARS-CoV-2 Spike and Nucleocapsid Proteins in Liquid Phase

Kai Wu[a,†,*], Vinit Kumar Chugh[a,†], Venkatramana D. Krishna[b,†], Arturo di Girolamo[a], Yongqiang Andrew Wang[c], Renata Saha[a], Shuang Liang[d], Maxim C-J Cheeran[b,*], and Jian-Ping Wang[a,*]

[a]Department of Electrical and Computer Engineering, University of Minnesota, Minneapolis, MN 55455, United States

[b]Department of Veterinary Population Medicine, University of Minnesota, St. Paul, MN 55108, United States

[c]Ocean Nano Tech LLC, San Diego, CA 92126, United States

[d]Department of Chemical Engineering and Material Science, University of Minnesota, Minneapolis, MN 55455, United States

[†]K.W., V.K.C., and V.D.K. have contributed equally to this work.

*E-mails: wuxx0803@umn.edu (K. W.); cheeran@umn.edu (M. C-J. C); jpwang@umn.edu (J.-P. W.)

## S1. Static magnetic hysteresis loops and magnetic properties of IPG30 MNPs.

The static magnetic hysteresis loops are collected from air-dried 10 μL IPG30 MNP and the magnetizations are calculated based on its weight concentration. As shown in Figure S1, magnetic fields are swept from -5000 Oe to +5000 Oe for Figure S1(a), and from -500 Oe to +500 Oe for Figure S1(b). The IPG30 MNPs show a saturation magnetization of 30.04 emu/g and specific magnetization of 22.3 emu/g at 500 Oe. The magnetic coercivity is around 15 Oe. Due to the surface coating layer of protein G, the inter-particle distance is increased thus, this small coercivity field does not cause the clustering or sedimentation of MNPs.

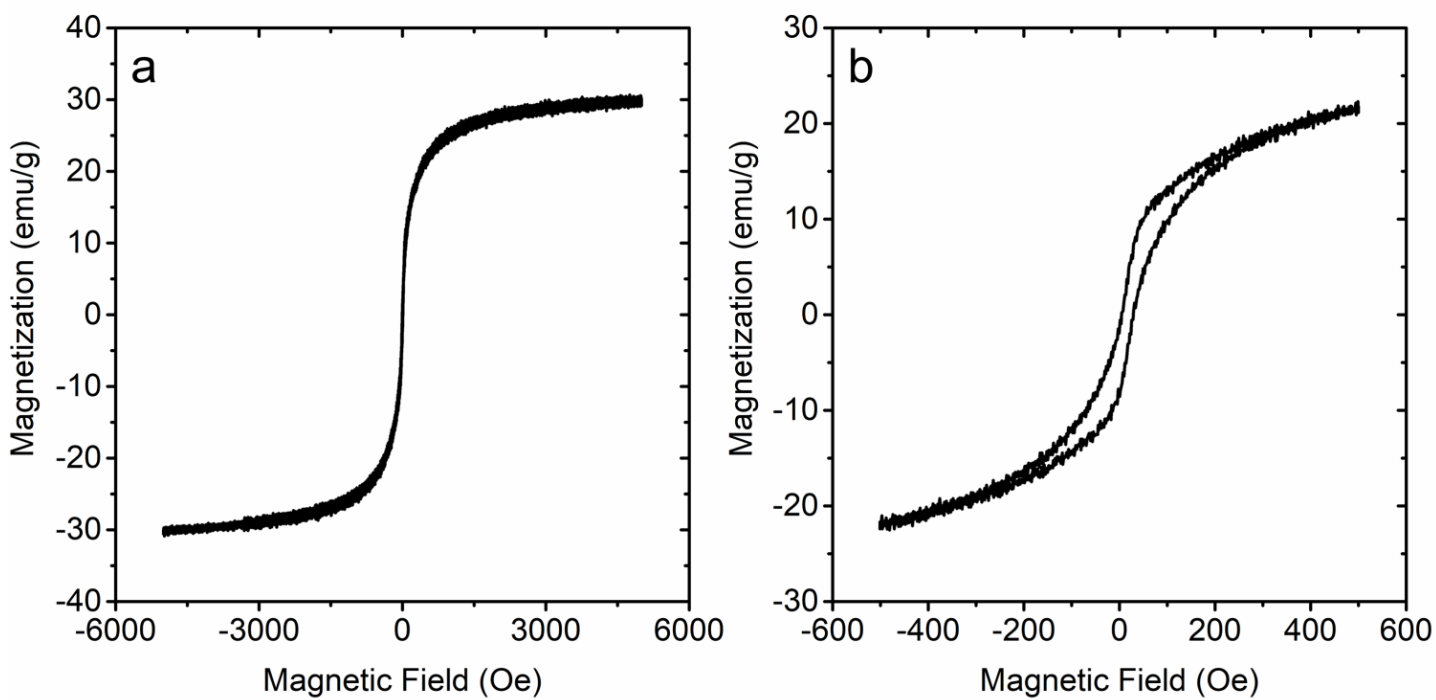

Figure S1. Static magnetic hysteresis loops of air-dried IPG30 MNPs measured by a PPMS. The field range is (a) -5000 – 5000 Oe and (b) -500 – 500 Oe, respectively.

**S2. The minimum detectable amount of IPG30 by homebuilt MPS system.**

The minimum detectable amount of IPG30 MNPs by our homebuilt MPS system is firstly investigated by two-fold dilutions of IPG30. There are a total of 13 vials prepared each contains 80 μL IPG30 of varying degrees of dilutions, and vial #13 is a blank sample for control purpose, as shown in Table S1. Three independent MPS readings are taken on each sample and the higher harmonics (the $3^{rd}$, $5^{th}$, $7^{th}$, and $9^{th}$) are summarized in Figure S2. Overall, the harmonic signal amplitudes decrease linearly as we decrease the amount of IPG30 MNPs in the vial. The figure insets zoom in the harmonic amplitudes of vials that contain less than 1 μg of IPG30 MNPs. All of the $3^{rd}$, the $5^{th}$, and the $7^{th}$ harmonics show the detection limit of IPG30 MNPs is at 265 ng (vial #10, 512-fold dilution), which equals to 9 fmole of IPG30 nanoparticles. In order to cut down the cost per assay and make MPS an inexpensive platform for high volume bioassays in the future, we explored the feasibility of using 20-fold diluted IPG30 for detection of SARS-CoV-2 nucleocapsid and spike proteins in this paper.

Table S1. IPG30 Samples Prepared by Two-fold Dilutions.

| Vial Index | IPG30 Concentration (mg/mL) | IPG30 Amount (μg) | Dilution (fold) |
| --- | --- | --- | --- |
| 1 | 1.7 | 136 | 1 |
| 2 | 0.85 | 68 | 2 |
| 3 | 0.425 | 34 | 4 |
| 4 | 0.2125 | 17 | 8 |
| 5 | 0.10625 | 8.5 | 16 |
| 6 | 0.053125 | 4.25 | 32 |
| 7 | 0.0265625 | 2.125 | 64 |
| 8 | 0.01328125 | 1.0625 | 128 |
| 9 | 0.006640625 | 0.53125 | 256 |
| 10 | 0.003320313 | 0.265625 | 512 |
| 11 | 0.001660156 | 0.1328125 | 1024 |
| 12 | 0.000830078 | 0.06640625 | 2048 |
| 13 | 0 | 0 | NA |

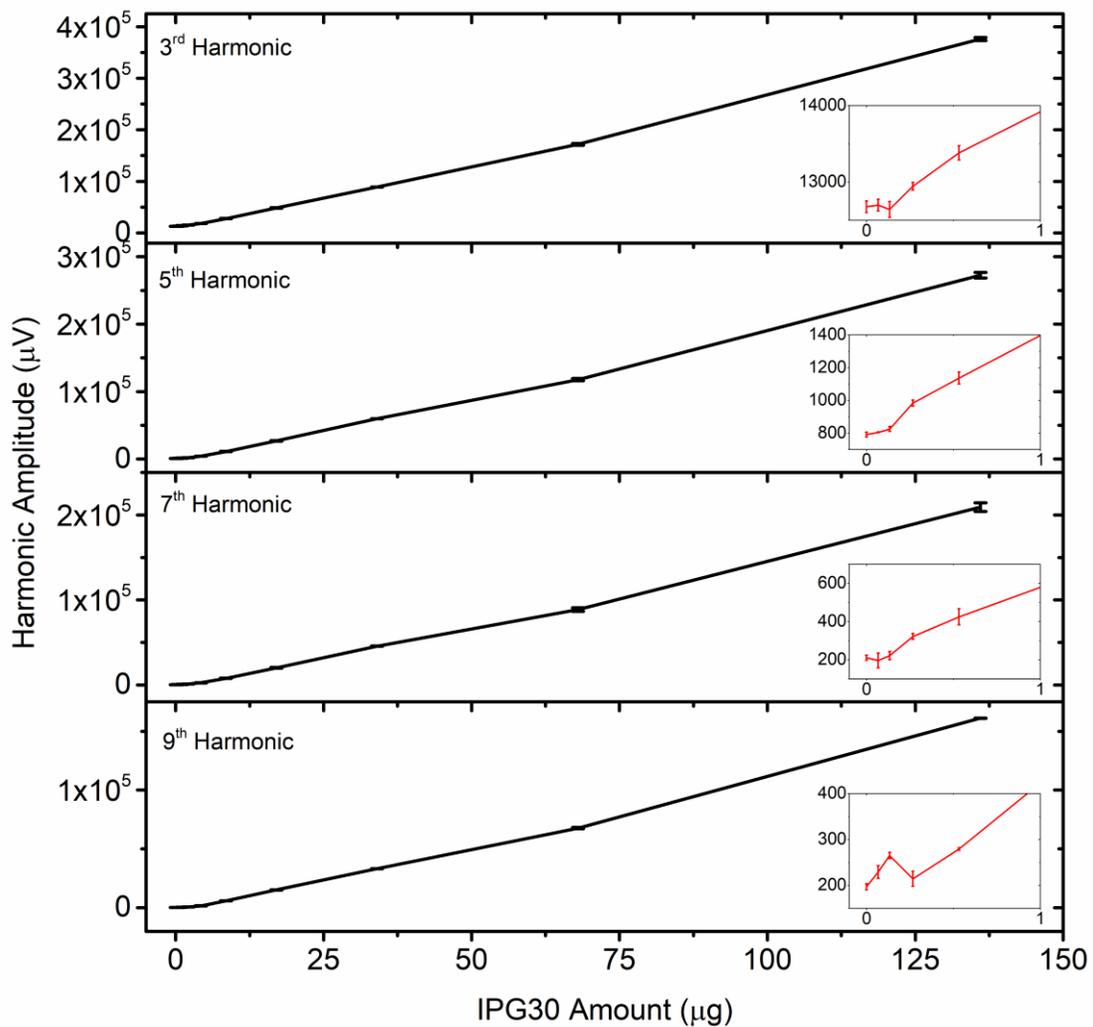

Figure S2. The higher harmonic amplitudes collected from vials 1-13 with varying amount of IPG30 MNPs. Error bars represent standard errors.

## S3. Hydrodynamic size, zeta potential, the 3rd harmonic amplitudes of bare MNPs and MNPs functionalized with different amount of pAbs.

To confirm the successful functionalization of different numbers of pAb on each MNP, the hydrodynamic sizes of MNPs are characterized by Dynamic Light Scatter (DLS, Microtrac NanoFlex). As shown in Figure S3, the bare MNPs show smallest averaged hydrodynamic size of 41.4 nm, followed by MNP:pAb=1:1 and MNP:pAb=1:2. With MNP:pAb=1:3 and MNP:pAb=1:4 showing largest hydrodynamic sizes.

The functionalization of different amount of pAbs will inevitably hinder the Brownian relaxation of MNPs, thus, with more pAbs functionalized on each MNP, the weaker magnetic responses (refelected in the MPS harmonic amplitudes) will be observed. As summarized in Table S2, the 3rd harmonic amplitudes of vials #9 (bare MNP), #IV-8 (MNP:pAb=1:4), #III-8 (MNP:pAb=1:3), #II-8 (MNP:pAb=1:2), and #I-8 (MNP:pAb=1:1) are 22.76 mV, 21.33 mV, 20.58 mV, 19.95 mV, and 18.18 mV, respectively.

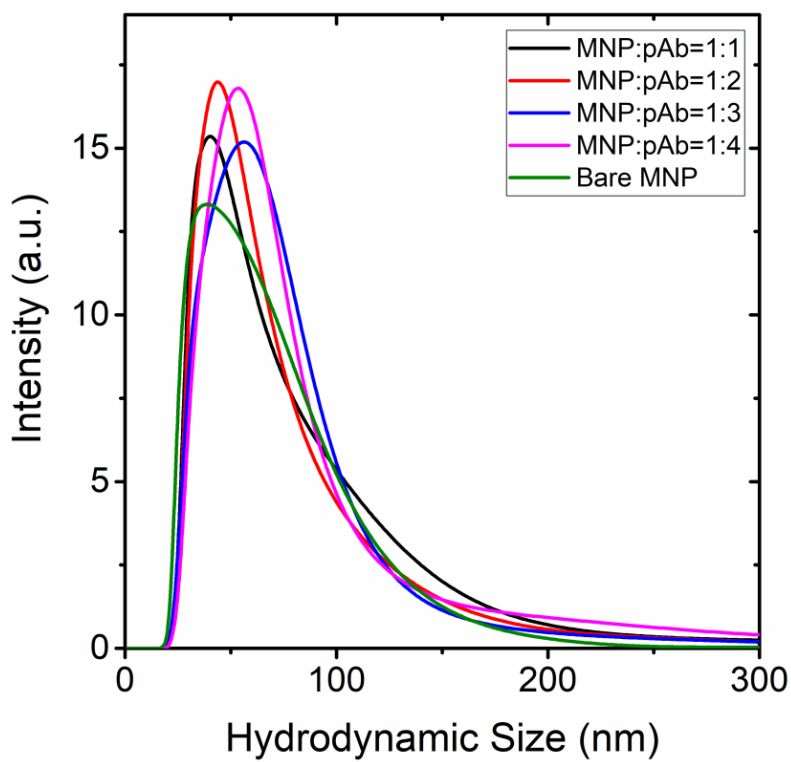

Figure S3. The hydrodynamic size distributions of bare MNPs, MNPs each functionalized with one, two, three, and four pAbs.

The colloidal stability of bare MNPs is compared with MNPs functionalized with pAbs. It is confirmed that the MNP suspensions show a neutral to slightly alkaline pH of 7.0 - 7.1. As shown in Table S2, with the functionalization of pAbs, the zeta potential barely changes. Thus, functionalizing pAbs to MNPs won't affect the stability of MNPs.

Table S2. Average hydrodynamic sizes, zeta potentials, and 3rd harmonic amplitudes of bare MNPs and MNPs functionalized with different amount of pAbs.

| Sample | Average Hydrodynamic Size (nm) | Zeta Potential (mV) | The Averaged 3rd Harmonic Amplitude (mV) |
|---|---|---|---|
| Bare MNP | 41.4 | -0.34 | 22.76 |
| MNP:pAb=1:1 | 44 | -0.33 | 21.33 |
| MNP:pAb=1:2 | 44.5 | -0.33 | 20.58 |
| MNP:pAb=1:3 | 46.4 | -0.34 | 19.95 |
| MNP:pAb=1:4 | 47.8 | -0.34 | 18.18 |

**S4. Grayscale heatmaps of higher harmonic signal drop Δ (in %) compared to bare MNPs for SARS-CoV-2 spike proteins.**

The 3rd, 5th, 7th, 9th, 11th, 13th, and 15th harmonics of each sample is compared with the corresponding harmonics from bare MNP sample (vial #9) and the grayscale heatmaps of harmonic signal drop (defined as $\Delta = \frac{Ai_9 - Ai_{X-j}}{Ai_9} \times 100\%$, where $i$ is the harmonic index, the subscripts are sample indexes, X=I, II, III, and IV, $j$=1, 2, 3, …, 8) are plotted. In each row of Figure S4(a – g), by adding the same amount of spike protein molecules, the harmonic signal drop Δ decreases from I to IV, which agrees with the results in Figure 4(e – i). Thus, ideally, the color becomes darker from left column to right column in each grayscale heatmap (as schemed in Figure S4(h)). In each column of Figure S4(a – g), with IPG30 MNPs surface functionalized with an identical number of pAbs, adding more spike protein causes larger harmonic signal drop. Thus, ideally, the color becomes darker from the bottom row (vials #X-1, X=I, II, III, and IV) to the top row (vials #X-8, X=I, II, III, and IV). Again, this ideal trend is schemed in Figure S4(h).

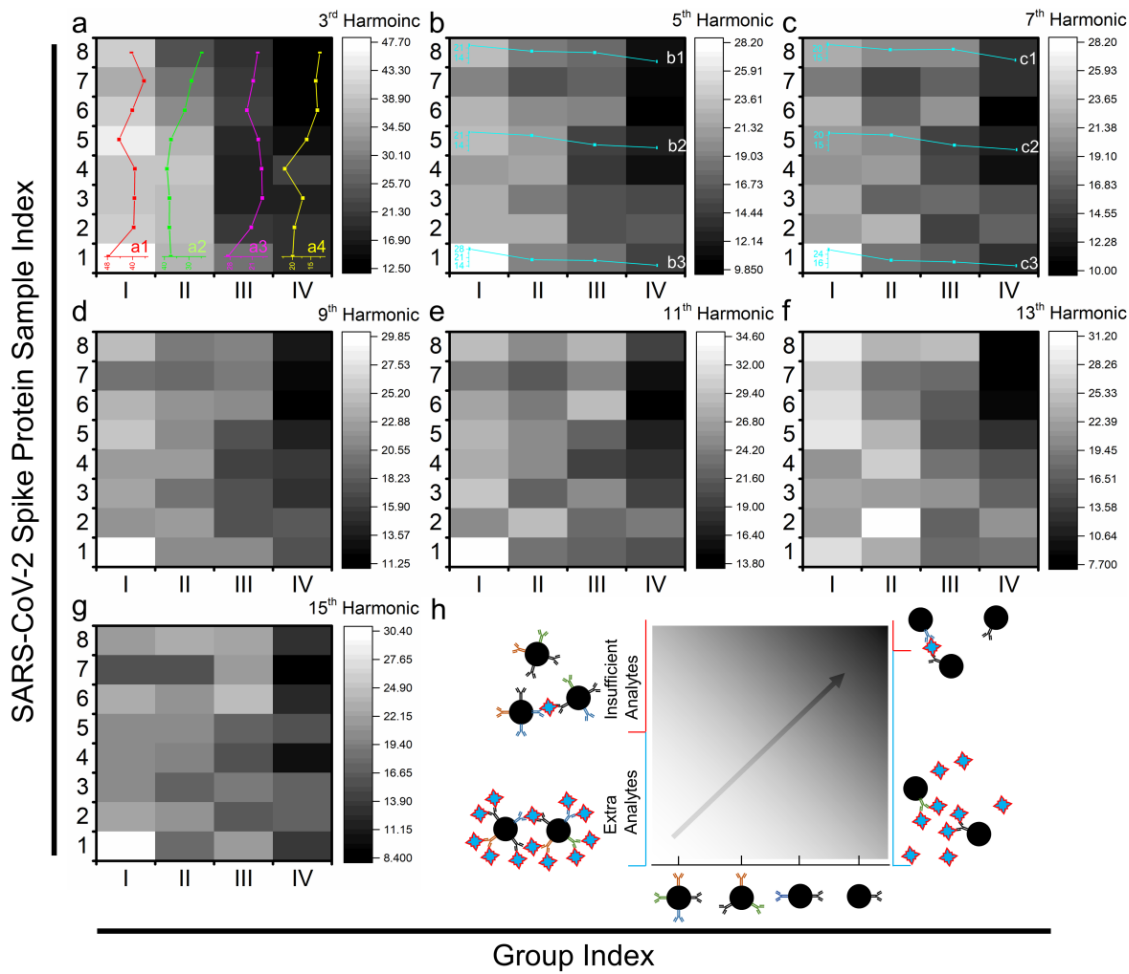

Figure S4. Grayscale heatmaps of higher harmonic signal drop Δ (in %) compared to bare MNPs for SARS-CoV-2 spike proteins. (a) – (g) are the 3rd, 5th, 7th, 9th, 11th, 13th, and 15th harmonic signal drop from 32 samples in Group I – IV compared to bare MNPs, respectively. (a1 - a4) are the harmonic signal drop, Δ, plotted as a function

S7

of spike protein amount/concentration for groups I – IV, respectively.  (h) is the grayscale heatmap showing the ideal color trend regarding different amount of pAbs functionalized on MNPs as well as different scenarios of extra and insufficient target analytes (i.e., SARS-CoV-2 spike protein).

The 3$^{rd}$ harmonics give highest signal to noise ratio (SNR), and we can clearly see this trend from Figure S4(a). Figure S4(a1 – a4) plots the 3$^{rd}$ harmonic signal drop Δ curves for samples from Groups I – IV, where Group II shows the best monotonic concentration-response curve. Figure S4(b: 1 – 3) and Figure 6(c: 1 – 3) plot the 5$^{th}$ and the 7$^{th}$ harmonic signal drop Δ curves for samples added with same amount of spike protein molecules, across Groups I – IV. Figure S4(h) schematically draws the scenarios where extra and insufficient target analytes (i.e., spike protein) are added. As a result, the number of target analytes directly affects the degree of nanoparticle clustering as well as the dynamic magnetic responses.

**S5. Grayscale heatmaps of higher harmonic signal drop Δ (in %) compared to bare MNPs for SARS-CoV-2 nucleocapsid proteins.**

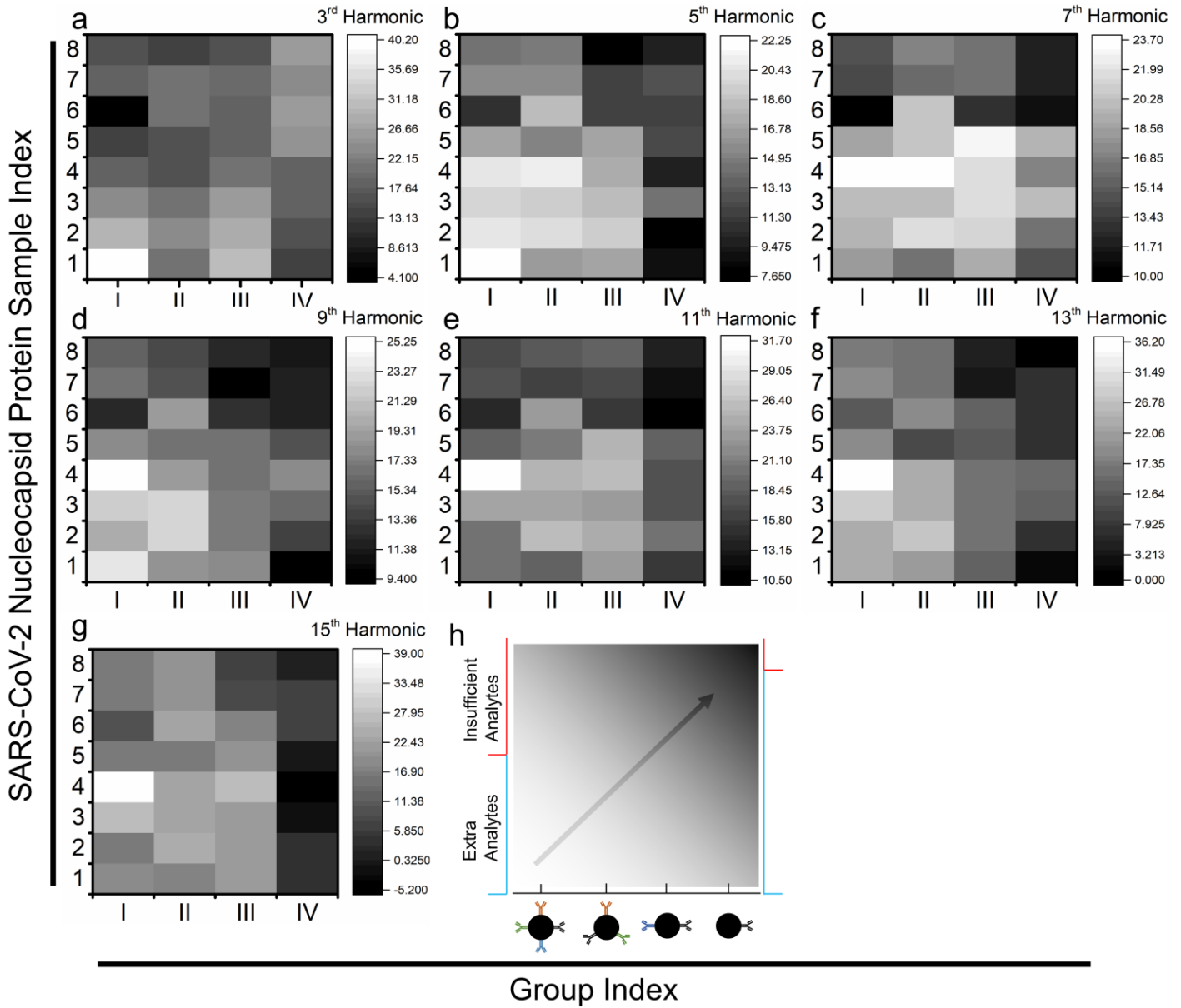

Figure S5. Grayscale heatmaps of higher harmonic signal drop Δ (in %) compared to bare MNPs for SARS-CoV-2 nucleocapsid proteins.

## S6. The ratios of higher harmonics to the 3rd harmonics recorded from groups I to IV for SARS-CoV-2 nucleocapsid protein.

The ratios of higher harmonics (from the 5th to the 15th harmonics) to the 3rd harmonics collected from each sample for SARS-CoV-2 nucleocapsid protein detection. As shown in Figure S6, for Group I, the harmonic ratio curves from samples #I-1 to #I-8 are sparsely distributed. The significant differences in harmonic ratio curves from vials added with different concentration/amount of nucleocapsid protein molecules allow us to analyze and collect meaningful concentration-response profiles as shown in Figure 7(a). However, for groups II and III, the harmonic ratio curves are tightly distributed with very narrow gaps or even overlapping. Indicating very small differences in harmonic signals for samples in Groups II and III, as shown in Figure 7(b) & (c). Interestingly, Group IV shows relatively sparse distributions of harmonic ratio curves compared with Groups II & III, while it's denser than Group I. The harmonic ratio curves from Group IV can be divided into three factions: vials#IV-1 to #IV-4 showing lower harmonic ratios are overlapping, vials #IV-6 to #IV-8 showing higher ratios are overlapping, and the harmonic ratio of vial#IV-5 (blue curve) is in between. Thus, Group IV is also not practical for the detection of nucleocapsid proteins.

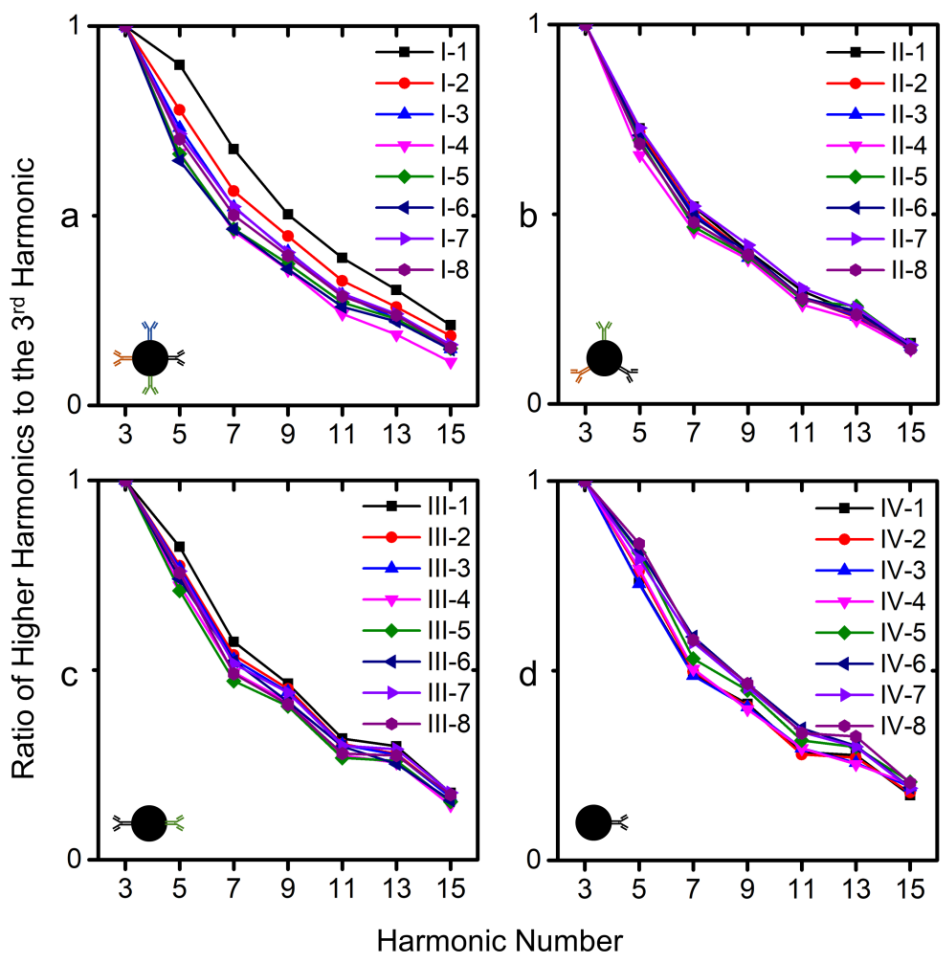

Figure S6. Ratios of higher harmonics to the 3rd harmonics recorded from Groups (a) I, (b) II, (c) III, and (d) IV. For SARS-CoV-2 nucleocapsid protein detection.

**S7. Specificity of SARS-CoV-2 spike and nucleocapsid pAbs confirmed by ELISA.**

As shown below, ELISA was performed using polyclonal antibody (pAbs) to spike or nucleocapsid as capture antibody and corresponding biotinylated pAbs as detection antibody. Microtiter wells were coated with Rabbit anti-Spike-RBD (for Spike ELISA) or rabbit anti-Nucleiocapsid (for nucleocapsid ELISA) as capture antibody and detected using corresponding biotin conjugated rabbit polyclonal antibodies and streptavidin-HRP. Different isolates of heat inactivated SARS-CoV-2 (USA-WA 1/2020, Hong Kong, Italy, USA-IL 1/2020), heat inactivated human corona viruses hCoV-NL63, hCoV-OC43, and hCoV-229E, porcine epidemic diarrhea virus (PEDV) and recombinant influenza A virus nucleoprotein (IAV NP) were tested using this ELISA. These pAbs to spike and nucleocapsid did not bind to other human corona viruses- hCoV-NL63, hCoV-OC43, and hCoV-229E and porcine epidemic diarrhea virus (PEDV), a corona virus causing diarrhea in pigs.

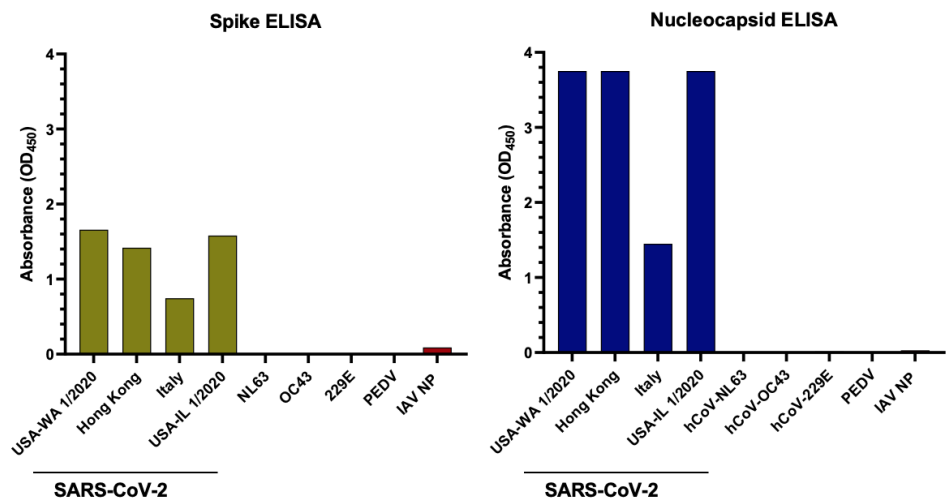

**Figure S7.** The binding specificities of pAbs are confirmed by testing four different SARS-CoV-2 isolates and four other coronaviruses using ELISA.